\documentclass[conference, 10pt, A4]{IEEEtran}
\IEEEoverridecommandlockouts

\usepackage{setspace}
\usepackage{float}
\usepackage{dblfloatfix}
\usepackage{cite}
\usepackage{amsmath,amssymb,amsfonts}
\usepackage{algorithm}
\usepackage{algpseudocode}
\usepackage{algorithmicx}
\usepackage{graphicx}
\usepackage{textcomp}
\usepackage{caption}
\usepackage{subcaption}
\usepackage{xcolor}

\usepackage{enumitem}
\usepackage[font=footnotesize]{caption}
\usepackage{tikz}

\newcommand*\circled[1]{\tikz[baseline=(char.base)]{
            \node[shape=circle,fill,inner sep=1pt] (char) {\textcolor{white}{#1}};}}

\makeatletter
\renewcommand\footnoterule{
  \kern-3\p@
  \hrule\@width 0.5\columnwidth
  \kern2.6\p@}
  \makeatother

\def\BibTeX{{\rm B\kern-.05em{\sc i\kern-.025em b}\kern-.08em
    T\kern-.1667em\lower.7ex\hbox{E}\kern-.125emX}}
    
\renewcommand{\IEEEbibitemsep}{0.3pt}
\makeatletter
\IEEEtriggercmd{\reset@font\normalfont\fontsize{7.2pt}{7.5pt}\selectfont}
\makeatother
\IEEEtriggeratref{1}

\begin{document}
\title{\textit{eFAT}: Improving the Effectiveness of Fault-Aware Training for Mitigating Permanent Faults in DNN Hardware Accelerators}

\author{\IEEEauthorblockN{Muhammad Abdullah Hanif, Muhammad Shafique}
\IEEEauthorblockA{\textit{Division of Engineering, New York University
Abu Dhabi (NYUAD), Abu Dhabi, United Arab Emirates}\\
mh6117@nyu.edu, muhammad.shafique@nyu.edu}}


\maketitle
\begin{abstract}
Due to the compute-intensive nature of Deep Neural Networks (DNNs), specialized hardware accelerators are employed to offer performance and energy-efficient inference at the edge. However, in the nanometer fabrication regime, these accelerators face various technology-induced reliability issues. One of the foremost reliability concerns is permanent faults induced due to imperfections in the chip fabrication process, as they can significantly reduce the manufacturing yield and thereby impact the cost of the marketed devices/systems. Fault-Aware Training (FAT) has emerged as a highly effective technique for addressing permanent faults in DNN accelerators, as it offers fault mitigation without significant performance or accuracy loss, specifically at low and moderate fault rates. However, it leads to very high retraining overheads, especially when used for large DNNs designed for complex AI applications. Moreover, as each fabricated chip can have a distinct fault pattern, FAT is required to be performed for each faulty chip individually, considering its unique fault map, which further aggravates the problem. To reduce the overheads of FAT while maintaining its benefits, we propose (1) the concepts of resilience-driven retraining amount selection, and (2) resilience-driven grouping and fusion of multiple fault maps (belonging to different chips) to perform consolidated retraining for a group of faulty chips. To realize these concepts, in this work, we present a novel framework, \textit{eFAT}, that computes the resilience of a given DNN to faults at different fault rates and with different levels of retraining, and it uses that knowledge to build a resilience map given a user-defined accuracy constraint. Then, it uses the resilience map to compute the amount of retraining required for each chip, considering its unique fault map. Afterward, it performs resilience and reward-driven grouping and fusion of fault maps to further reduce the number of retraining iterations required for tuning the given DNN for the given set of faulty chips. We demonstrate the effectiveness of our framework for a systolic array-based DNN accelerator experiencing permanent faults in the computational array. Our extensive results for numerous chips show that the proposed technique significantly reduces the retraining cost when used for tuning a DNN for multiple faulty chips. 

\end{abstract}
\section{Introduction}
\label{Sec1:Introduction}

Deep Neural Networks (DNNs) have demonstrated remarkable potential for solving challenging Machine Learning (ML)/Artificial Intelligence (AI) problems such as image classification, object detection, video processing, event recognition, scene classification, and language translation~\cite{lecun2015deep, sze2017efficient, pouyanfar2018survey}. However, state-of-the-art DNNs are highly compute-intensive and require a lot of resources to execute. To offer performance and energy-efficient inference at the edge, specialized DNN hardware accelerators, such as Eyeriss~\cite{chen2016eyeriss}\cite{chen2019eyeriss}, Bit~Fusion~\cite{sharma2018bit}, and TPU~\cite{jouppi2017datacenter}, are designed and deployed in advanced AI/ML systems. These accelerators are usually fabricated using nano-scale CMOS technologies, which experiences diverse reliability threats. 

One of the key reliability threats faced by nano-scale CMOS devices is permanent faults induced due to imperfections in the chip fabrication process. Prior works, such as~\cite{zhang2018analyzing}, have shown that even a small number of these faults can severely impact the accuracy of DNNs.
Hence, all the faulty chips must be discarded or go through post-fabrication optimizations, which drastically reduces the manufacturing yield and/or increases the cost of the marketed (fault-free) devices. 

\textbf{State-of-the-art and their Limitations:} To address permanent faults in DNN accelerators, various fault-mitigation techniques have been proposed in the literature. 
These techniques can be divided into three main categories. 
(1) The first category includes techniques that offer fault mitigation at the cost of performance loss by employing \textit{redundancy or disconnection of faulty elements}. 
For example, Kim et al.~\cite{kim1989design} proposed bypassing faulty Processing Elements (PEs) and viewing a faulty array as a smaller fault-free array with fewer rows/columns. Takanami et al.~\cite{takanami2012built}\cite{takanami2017built} proposed adding \textit{redundant PEs} in the architecture to handle faults in the computational array of the hardware. (2) The second category includes techniques that \textit{exploit the intrinsic fault tolerance of DNNs} to achieve low-cost fault mitigation. 
For example, \textit{Fault-Aware Pruning (FAP)}~\cite{zhang2018analyzing} exploits the resilience of DNNs to pruning to address permanent faults in the computational array of a systolic array-based DNN accelerator. \textit{Fault-Aware Mapping (FAM)}~\cite{abdullah2020salvagednn} improves \textit{FAP} by mapping less significant weights to the faulty (bypassed/zeroed) PEs. 
The main drawback of \textit{FAP} and \textit{FAM} is that these techniques result in accuracy loss, which is somewhat dependent on the number of faulty elements in the array. (3) The third category includes techniques that exploit fault-aware retraining to generate fault-aware DNNs. For example, \textit{Fault-Aware Pruning + Training (FAP+T)} is proposed in~\cite{zhang2018analyzing} to offer fault-mitigation with minimal performance and accuracy loss. 
\textit{Fault-Aware Training (FAT)} is also exploited in~\cite{zhang2019fault}\cite{MATIC_J}\cite{xu2019resilient} to mitigate permanent and other types of faults in DNN accelerators. 
The above works clearly show that \textit{FAT} leads to the best accuracy results. However, \textit{the main drawback of such techniques is that they incur huge retraining overheads, specifically when a DNN has to be tuned for a large number of faulty chips having distinct fault maps (detailed motivational case study is presented in Section~\ref{sec:motivational_case_study}).}
Works like~\cite{hoang2021tre} have tried addressing the overheads of FAT by merging multiple fault maps belonging to different faulty chips; however, they show that a reduction in the retraining overheads can only be achieved at the cost of accuracy loss. 

\textbf{Targeted Research Problem:} In this work, we aim to address the challenging question of how to reduce the (re)training overheads of FAT for cases where a given DNN has to be tuned for a large number of faulty chips having different fault maps. 
In order to illustrate the impact of DNN complexity and the number of faulty chips on the overall FAT cost and to highlight the relationship between the amount of retraining required to reach a user-defined accuracy constraint and the fault characteristics of the DNN hardware, we present two case studies in the following subsection.  

\subsection{Motivational Case Studies and Research Challenges}
\label{sec:motivational_case_study}

\begin{figure}[tb]
\centering
\includegraphics[width=1\linewidth]{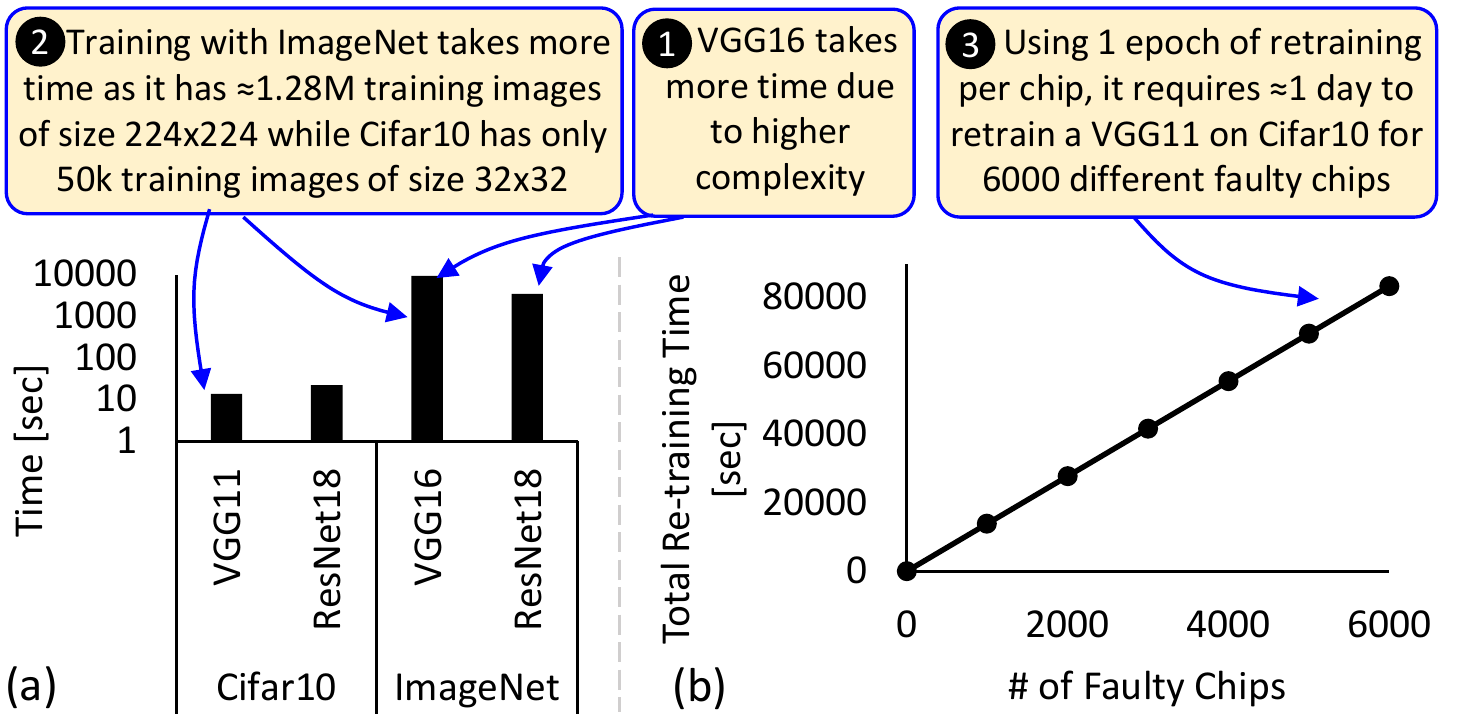}
\caption{(a) Execution time of one epoch of retraining for different DNN and dataset pairs. (b) Overall retraining time for tuning VGG11 on Cifar10 dataset for multiple faulty chips using 1 epoch of retraining per chip.}
\label{fig:Motivational_Figure_1}
\end{figure}

There are several factors that drive the overall retraining cost.
Fig.~\ref{fig:Motivational_Figure_1}a presents the execution time of one~epoch of training for different DNN and dataset pairs. As can be seen in the figure, the execution time varies significantly depending on the size and complexity of both the DNN and the dataset (see labels \circled{1} and \circled{2} in Fig.~\ref{fig:Motivational_Figure_1}). 
For example, it takes $\approx$13.9~sec to complete one~epoch of retraining of VGG11 on Cifar10 dataset, while $\approx$150~mins are required to complete one~epoch of retraining of VGG16 on ImageNet dataset using a high-end machine equipped with one RTX~3080 GPU\footnote{Detailed experimental setup is presented in Section~\ref{sec:experimental_setup}}. 
Apart from the DNN and the dataset, the number of faulty chips for which a DNN has to be retrained also affects the overall retraining cost. 
Fig.~\ref{fig:Motivational_Figure_1}b shows that the cost of retraining increases linearly with the increase in the number of faulty chips, assuming the same amount of retraining is performed for each faulty chip. 
It also highlights that even for relatively less complex DNN and dataset pairs (e.g., in this case VGG11 and Cifar10), as the number of faulty chips increases, the overall retraining cost can reach undesirable limits (see label \circled{3} in Fig.~\ref{fig:Motivational_Figure_1}). \textit{In summary, Fig.~\ref{fig:Motivational_Figure_1} shows that retraining a DNN for multiple faulty chips can be extremely costly depending on the complexity of the DNN, the dataset and the number of faulty chips.}

\begin{figure}[ht]
\centering
\includegraphics[width=1\linewidth]{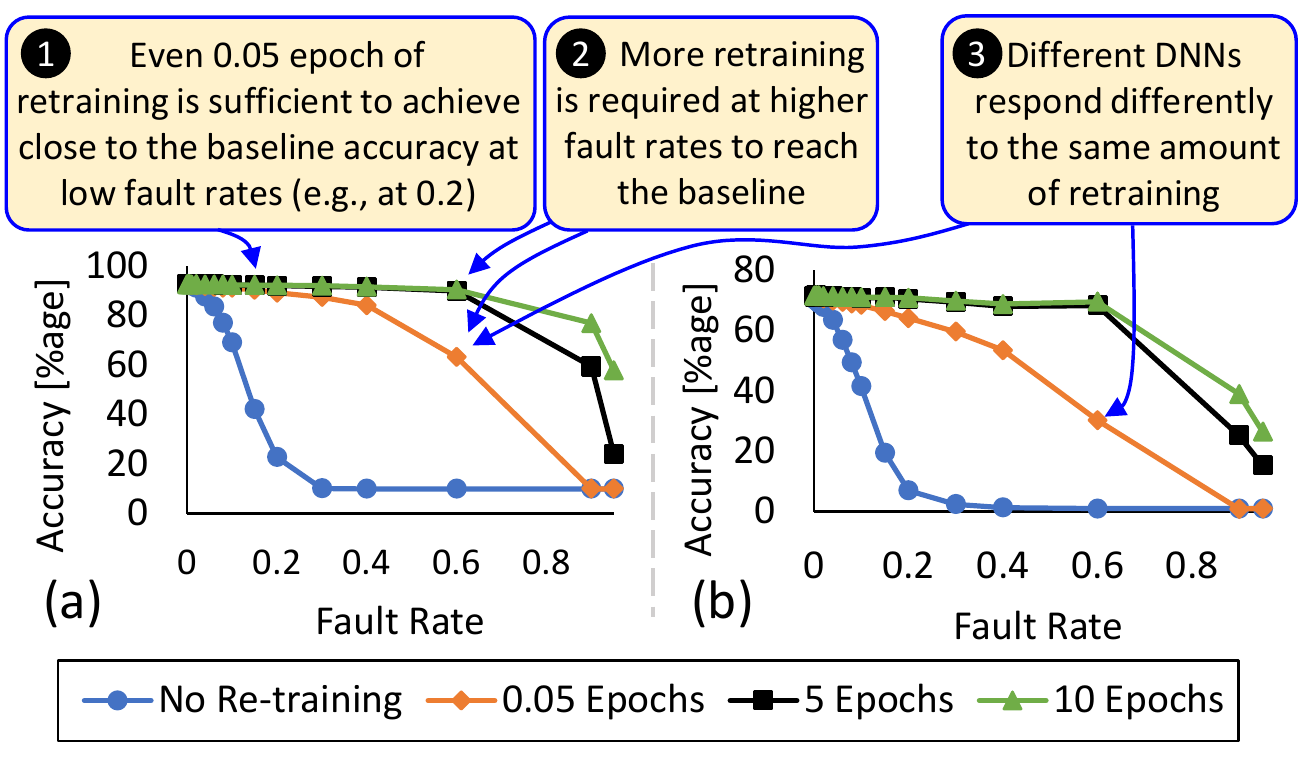}
\caption{\textbf{(a)} Accuracy results of \textbf{VGG11 trained on Cifar10 dataset} at different fault rates and with different levels of retraining. \textbf{(b)} Accuracy results of \textbf{ResNet18 trained on Cifar100 dataset}. Note, the fault rate in these examples represent the ratio of the number of faulty Processing Elements (PEs) to the number of total PEs in the computational array.}
\label{fig:Motivational_Figure_2}
\end{figure}

While we assume that the pre-trained DNN is provided as an input to the methodology by the user, we cannot modify the structure of the DNN to reduce the retraining cost. 
However, we can define the amount of retraining required for each individual chip based on its characteristics such as the number and types of faults. 
To understand this, consider the case of permanent faults in the computational array of a DNN accelerator together with the fault mitigation technique \textit{FAP+T} from~\cite{zhang2018analyzing}. 
Fig.~\ref{fig:Motivational_Figure_2} shows the accuracy results of two different DNNs, VGG11 trained on Cifar10 dataset and ResNet18 trained on Cifar100 dataset, at different fault rates and with different levels of retraining. The results show that at low fault rates even very small amount of retraining is sufficient to regain the lost accuracy (see label \circled{1} in Fig.~\ref{fig:Motivational_Figure_2}), and at higher fault rates, more retaining is required to achieve close to the baseline results (see label \circled{2} in Fig.~\ref{fig:Motivational_Figure_2}). Moreover, different DNNs have different resilience to faults and respond differently to the same amount of retraining at same fault rates (see label \circled{3} in Fig.~\ref{fig:Motivational_Figure_2}). \textit{This analysis points towards the need for an adaptive approach that defines the amount of retraining required for each chip based on its fault characteristics and the resilience of the given DNN to faults.} 

\begin{figure}[bt]
\centering
\includegraphics[width=1\linewidth]{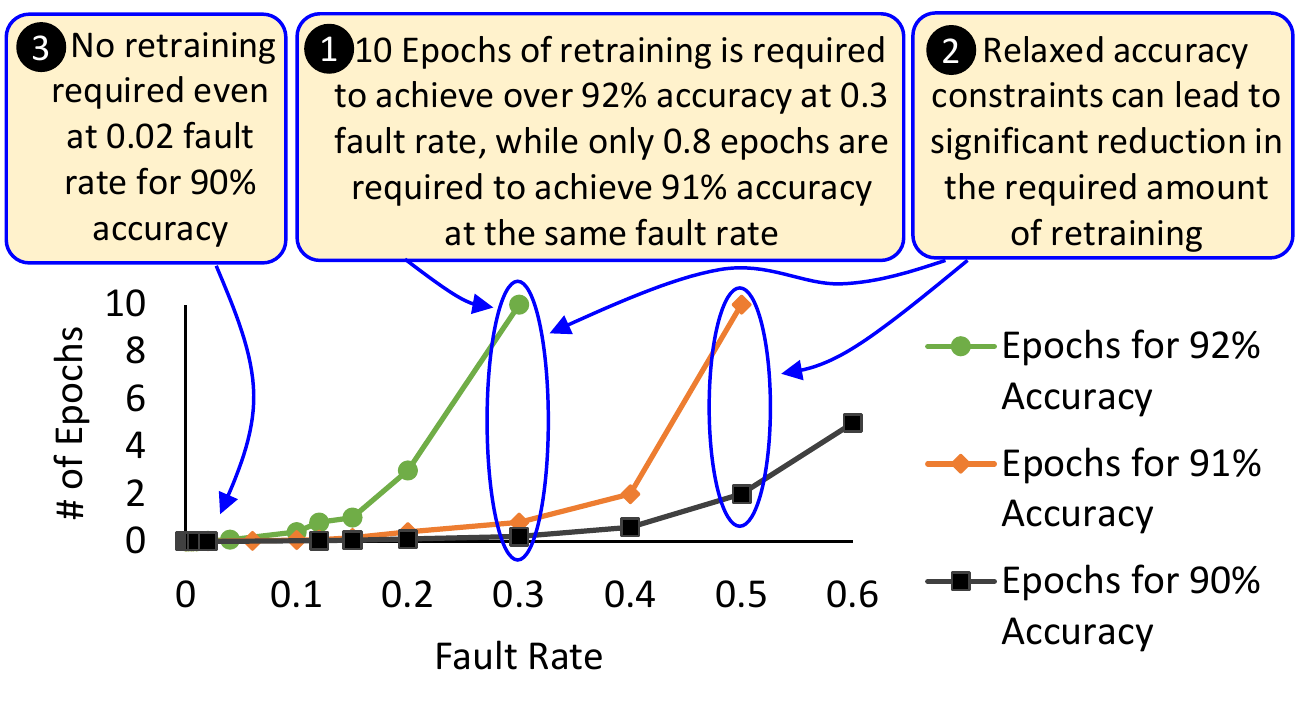}
\caption{Number of retraining epochs required by VGG11 to reach a pre-defined accuracy constraint on Cifar10 dataset at different fault rates. }
\label{fig:Motivational_Figure_3}
\end{figure}

\begin{figure*}[hbt]
\centering
\includegraphics[width=1\linewidth]{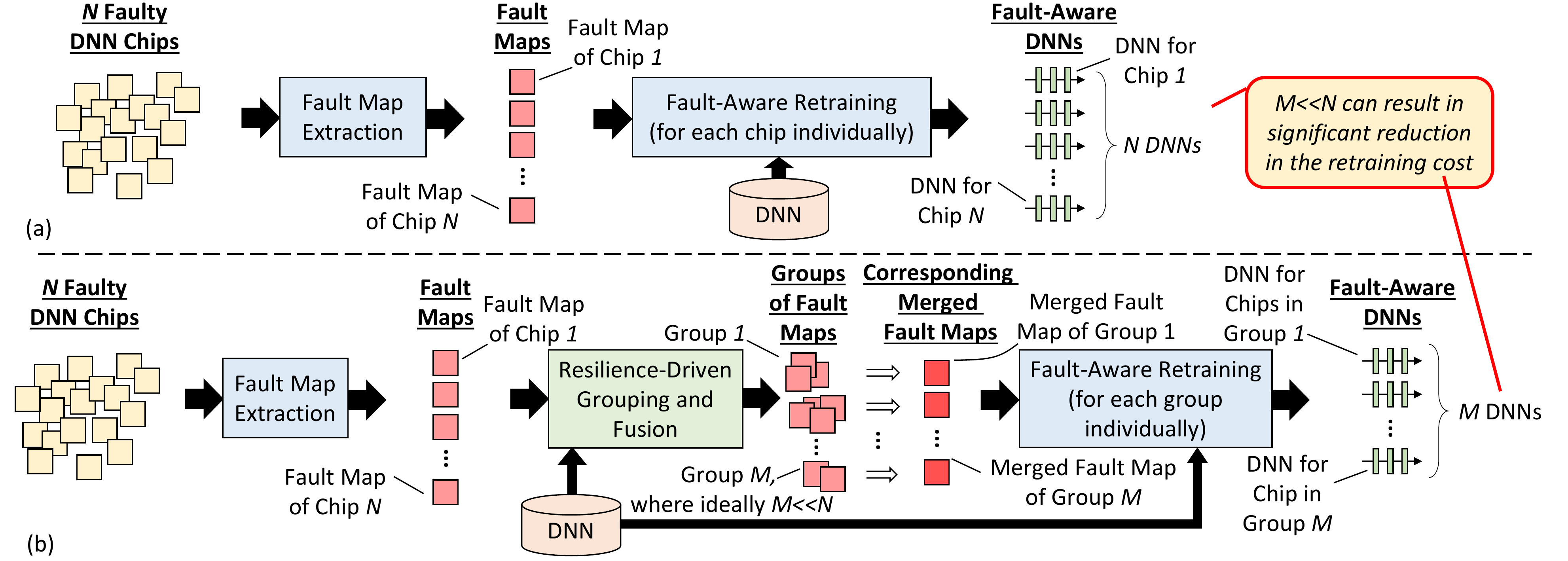}
\caption{(a) Conventional method of performing FAT for multiple faulty chips. (b) Our novel concept of resilience-driven grouping and fusion of fault maps to reduce the cost of FAT when used for tuning a DNN for multiple faulty chips.}
\label{fig:Motivation_4}
\end{figure*}

The amount of retraining required does not depend only on the number and types of faults in the chip. It also depends on the user-defined accuracy constraint. Fig.~\ref{fig:Motivational_Figure_3} shows this for the VGG11 trained on Cifar10 dataset. The figure shows that different accuracy constraints lead to different retraining requirements, and in the case of slightly relaxed accuracy constraints, significant efficiency gains can be achieved through intelligent retraining amount selection for each individual faulty chip based on its fault characteristics (see labels \circled{1} and \circled{2} in Fig.~\ref{fig:Motivational_Figure_3}). 
Moreover, relaxed constraints may lead to no need for retraining at lower fault rates (see label \circled{3} in Fig.~\ref{fig:Motivational_Figure_3}). 

Apart from the amount of retraining for each individual chip, we can also target the number of retraining iterations to reduce the overall retraining cost for multiple faulty chips. Fig.~\ref{fig:Motivation_4}a shows the conventional method where FAT is performed for each chip individually. 
However, based on the correlation between different fault maps and the resilience characteristics of the given DNN, resilience-driven grouping and fusion of fault maps (see Fig.~\ref{fig:Motivation_4}b) can be exploited to reduce the average (per chip) retraining cost. 

\textbf{Summary of Research Challenges:} Exploiting the above concepts requires to address the following challenging research questions. 

\begin{itemize}
    \item How to determine an appropriate amount of retraining required for each individual chip based on its fault characteristics and user-defined constraints?
    \item How to determine if merging multiple fault maps would lead to any benefits (i.e., reduction in the overall retraining cost)?
    \item If merging fault maps can lead to benefits, how to decide which fault maps should be fused together? 
\end{itemize}

\subsection{Our Novel Contributions}

To address the above-mentioned research questions and reduce the overall retraining overheads of FAT when used for tuning a DNN for multiple faulty chips, in this work, we present:
\begin{enumerate}
    \item A novel framework, \textit{eFAT}, which estimates the resilience of the given DNN to faults, and defines the amount of retraining required for each individual faulty chip based on its fault characteristics and the resilience characteristics of the DNN. 
    \item A resilience-driven strategy for estimating the amount of retraining for each individual faulty chip. 
    \item A resilience-driven grouping and fusion strategy for deciding if, when and which fault maps should be merged together to reduce the total number of retraining iterations.
    \item Results to show the effectiveness of the proposed framework under different scenarios. 
\end{enumerate}

\textbf{Paper Organization:} Section~\ref{Sec:Background_} presents the necessary background. Section~\ref{Sec:proposed_framework} presents our overall framework in detail. Section~\ref{Sec:results} presents the results and discussion, followed by the conclusion in Section~\ref{Sec:conclusion}.

\section{Background and Related Works}
\label{Sec:Background_}
\subsection{Deep Neural Networks (DNNs)}

A DNN is a network of neurons that are arranged in the form of layers. A neuron is the fundamental unit of a neural network, which performs a weighted sum operation and then passes the resultant value through a non-linear activation function. The functionality of a neuron is mathematically represented as: 

\begin{figure}[ht]
\centering
\includegraphics[width=1\linewidth]{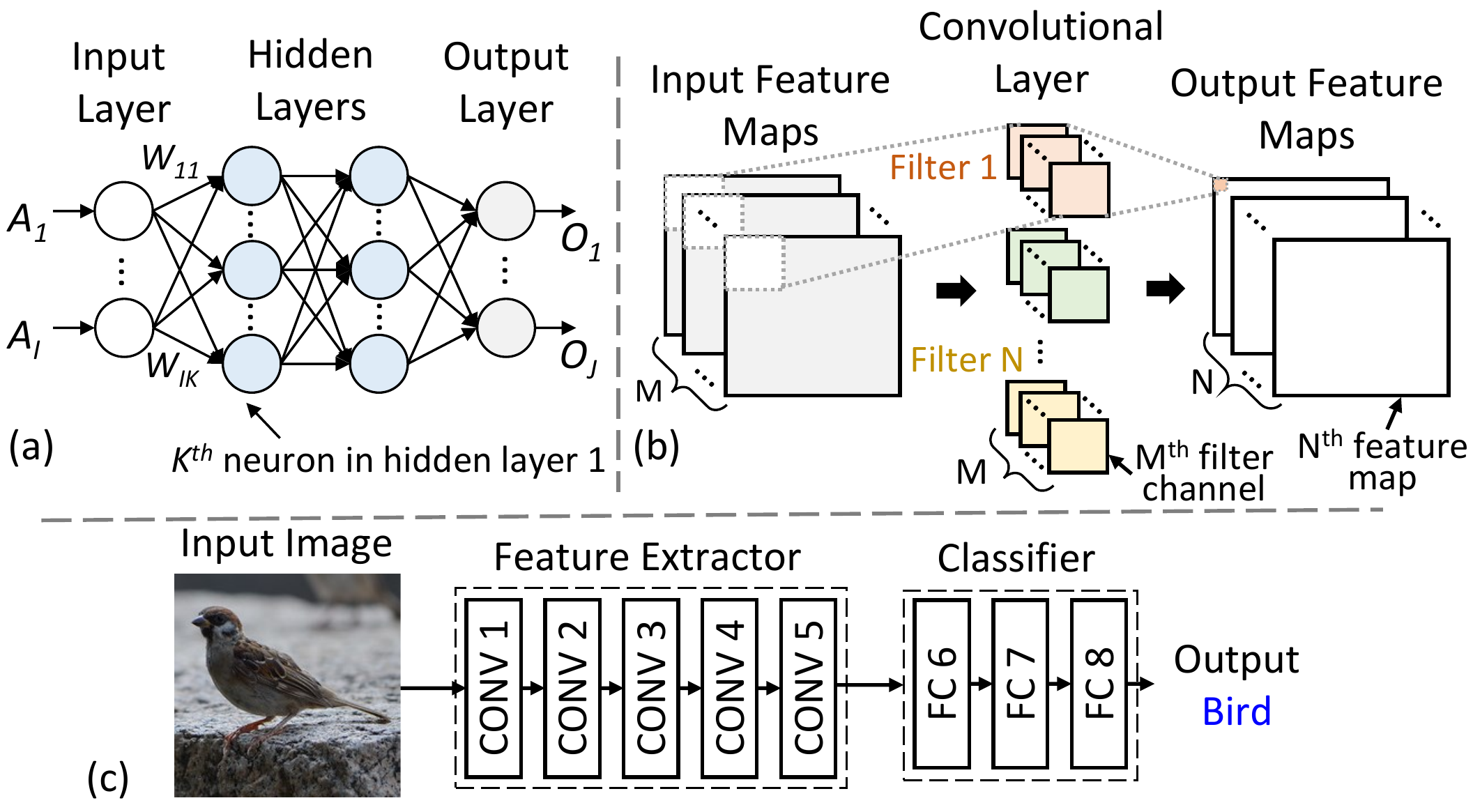}
\caption{(a) Example of a FC network. (b) A detailed view of a convolutional layer. (c) Example of a CNN.}
\label{fig:nn1}
\end{figure}

\begin{figure*}[h]
\centering
\includegraphics[width=1\linewidth]{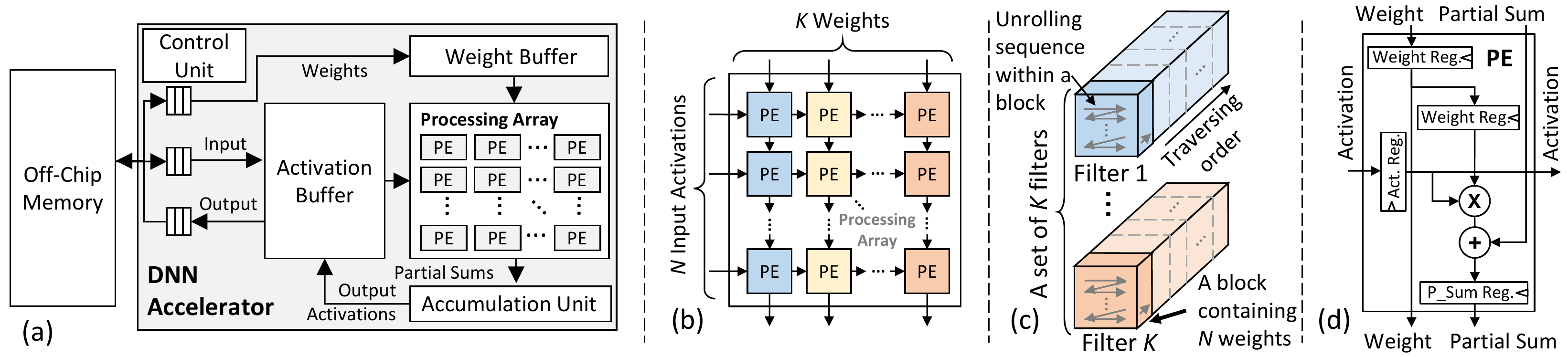}
\caption{(a) Overview of a DNN hardware accelerator. (b) A detailed view of a TPU-like systolic array. The shades of the PEs together with (c) show the mapping policy adopted for such arrays. (d) A detailed view of the PE architecture used in (b).}
\label{fig:hardware}
\end{figure*}

\begin{equation}
    Out = f(b + \sum_i W_{i} \times A_{i})
\end{equation}

Here, $Out$ represents the output, $A_i$ represents the $i^{th}$ input (a.k.a activation), $W_i$ represents the weight of the connection to the $i^{th}$ input, $b$ represents the bias, and $f(.)$ represents the non-linear activation function. 
Fig.~\ref{fig:nn1}a shows how these neurons are connected together to form a Fully-Connected (FC) network.  
Convolutional Neural Networks (CNNs) are another type of DNN used for processing images and videos. 
Fig.~\ref{fig:nn1}b shows a detailed view of a convolutional layer having $N$ number of filters, and Fig.~\ref{fig:nn1}c shows how these convolutional and FC layers can be connected to form a DNN for image classification application. More details on different types of neural networks can be found in~\cite{lecun2015deep}. 

\subsection{DNN Accelerators and Dataflow}

A significant amount of research have been carried out in designing specialized hardware accelerators, specifically for improving the efficiency of DNN inference process~\cite{jouppi2017datacenter, chen2019eyeriss, sharma2018bit, Myriad}.
As shown in the Fig.~\ref{fig:hardware}a, a DNN accelerator is mainly composed of a processing array and on-chip buffers. 
Fig.~\ref{fig:hardware}b shows a detailed view of a processing array similar to the one used in the Tensor Processing Unit (TPU)~\cite{jouppi2017datacenter}. 
For performing matrix/vector multiplications, the array is first loaded with weights through vertical channels. 
The weights are then kept stationary inside the array while the activations are fed through the horizontal channels. 
The partial sums flow downstream where they are added to the corresponding partial products. 
Fig.~\ref{fig:hardware}d shows the detailed architecture of the Processing Elements (PEs) used in the array, and Fig.~\ref{fig:hardware}c together with Fig.~\ref{fig:hardware}b illustrates the weight-stationary dataflow adopted for processing convolutional as well as FC layers. 
As shown in Fig.~\ref{fig:hardware}c, larger filters are divided into segments, which are then mapped one-by-one to the array to perform corresponding computations. 
Note that as partial sums flow downstream, only the weights belonging to the same filter/neuron can be mapped to the same column, while weights belonging to a different filter/neuron (but from the same layer and corresponding locations) can be mapped to different columns simultaneously. 

\section{\textit{eFAT}: Proposed Framework for Effective Fault-Aware Retraining}
\label{Sec:proposed_framework}

\subsection{Overview}
\begin{figure*}[htb]
\centering
\includegraphics[width=1\linewidth]{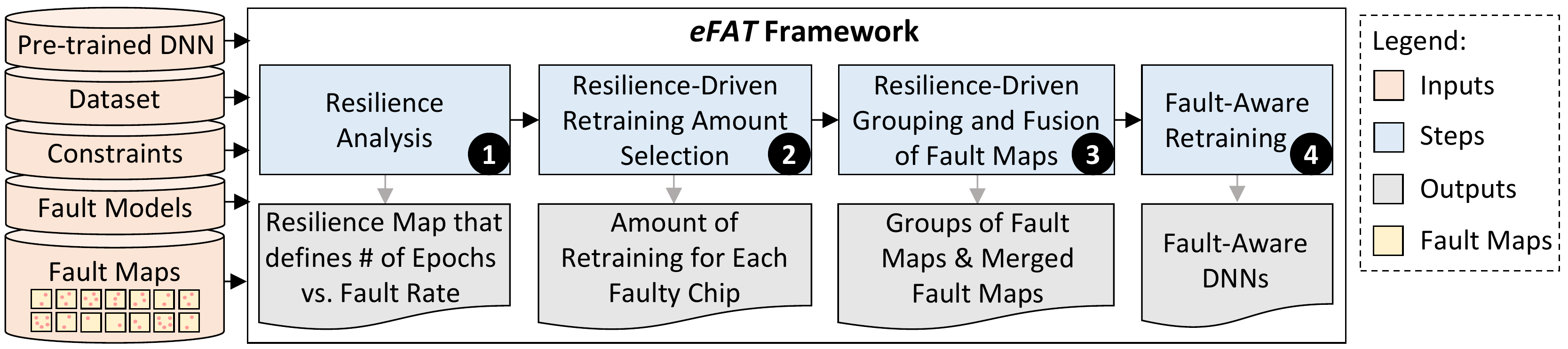}
\caption{Overview of the proposed \textit{eFAT} framework}
\label{fig:overall_framework}
\end{figure*}

Fig.~\ref{fig:overall_framework} shows the overview of our proposed \textit{eFAT} framework, which receives a pre-trained DNN, a training dataset, a user-defined accuracy constraint, and fault maps of the faulty chips as input. It defines a retraining policy to efficiently generate fault-aware DNNs for the given faulty chips. 
The framework first computes the resilience of the given DNN to faults using fault-injection experiments at different fault rates and with different levels of retraining (Step~\circled{1}). 
This resilience is then used in Step~\circled{2} to select the amount of fault-aware retraining for each individual faulty chip based on its fault characteristics. 
The selection is performed in such a way that guarantees the user-defined accuracy constraint is met without incurring unnecessary overheads. 
Step~\circled{3} then uses the selected values together with the resilience of the DNN to identify if and which fault maps can be merged together to further reduce the overall retraining cost. 
This step performs resilience-driven grouping and fusion of fault maps, and generates the final set of fault maps for which \textit{FAT} has to be performed. 
It also defines the amount of retraining for each merged fault map. 
In the final step, Step~\circled{4}, \textit{FAT} is performed and the generated fault-aware DNNs are then distributed to their corresponding faulty chips. The following subsections explain these different steps of our eFAT framework in detail. 

\subsection{Resilience Analysis (Step 1)}

\begin{figure*}[ht]
\centering
\includegraphics[width=1\linewidth]{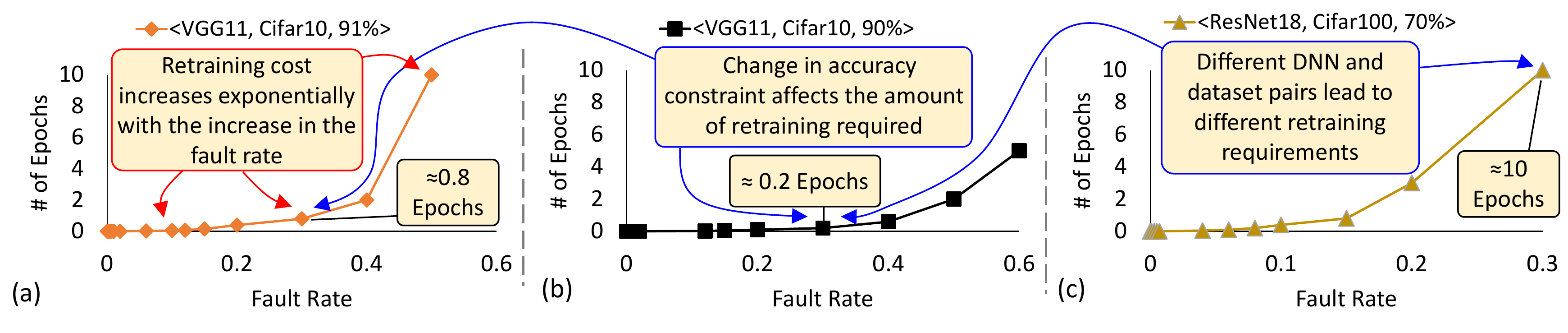}
\caption{Amount of fault-aware retraining required for the VGG11 (trained on Cifar10 dataset) at different fault rates to reach the user-defined accuracy constraint of 91\% (a) and 90\% (b). (c) Amount of fault-aware retraining required for the ResNet18 (trained on Cifar100 dataset) at different fault rates to reach the user-defined accuracy constraint of 70\%. Note that these results are generated assuming the case where faults occur in the computational array of a DNN accelerator and the hardware architecture is equipped with the additional circuitry to support FAP.} 
\label{fig:Resilience_Results} 
\end{figure*}

As highlighted in the introduction using Figs.~\ref{fig:Motivational_Figure_2} and~\ref{fig:Motivational_Figure_3}, resilience of DNNs to faults depends on various factors. 
The resilience indirectly affects the amount of retraining required to meet the user-defined accuracy constraint. 
In this step, we mainly compute the amount of retraining required at different fault rates to reach the user-defined accuracy level. 
We achieve this by performing FAT experiments using the provided fault models for fault injection. 
Starting from the minimum fault rate observed in the input fault maps, we compute the required amount of retraining at all the fault rates generated by Algo.~\ref{Algo:1}. 
Fig.~\ref{fig:Resilience_Results} presents example plots representing required amount of retraining at different fault rates. 
Note that each data point in these plots is generated by averaging the results over multiple iterations to cope with the variations in fault patterns across experiments. 

\begin{algorithm}[h]
\scriptsize	
\caption{List of fault rates for the \textit{required amount of retraining vs. fault rate} plot}
\label{Algo:1}
\begin{algorithmic}[1]
\Statex \textbf{Inputs:} Fault maps of the faulty chips; Maximum fault rate limit ($Max\_FR$); Maximum fault rate interval size ($Max\_int$); Step ratio ($Step$)
\Statex \textbf{Outputs:} A list of fault rates (\textbf{\textit{LFR}})
\Statex \textbf{Initialize:} \textbf{\textit{LFR}} = [ ], $Current\_FR = 0$ 
\State \textbf{\textit{FRs}} = Compute fault rates of the given fault maps
\State $Current\_FR = $ Min(\textbf{\textit{FRs}}) 
\State \textbf{\textit{LFR}}.append($Current\_FR$)
\While{$Current\_FR \leq $ Max([\textbf{\textit{FRs}}, $Max\_FR$])}
\State $Current\_FR$ += Min($Current\_FR * Step$, $Max\_int$) 
\State \textbf{\textit{LFR}}.append($Current\_FR$)
\EndWhile
\State \Return \textbf{\textit{LFR}}
\end{algorithmic}
\end{algorithm}

Unlike the examples presented in Fig.~\ref{fig:Resilience_Results} in which only a single type of fault can occur, if multiple types of faults can occur (e.g., stuck-at-0 and stuck-at-1 in the on-chip weight memory) in a system, a multi-dimensional plot that presents the required amount of retraining at different combinations of fault rates maybe required. 
Moreover, additional plots generated using higher accuracy constraints as well as the accuracy plot without any fault-aware retraining may also be required, specifically for cases where multiple fault maps can be merged and the resultant fault map has contradicting faults (i.e., locations not having just a single type of fault across given fault maps). 

\subsection{Resilience-Driven Retraining Amount Selection (Step~2)}

\begin{figure}[htb]
\centering
\includegraphics[width=1\linewidth]{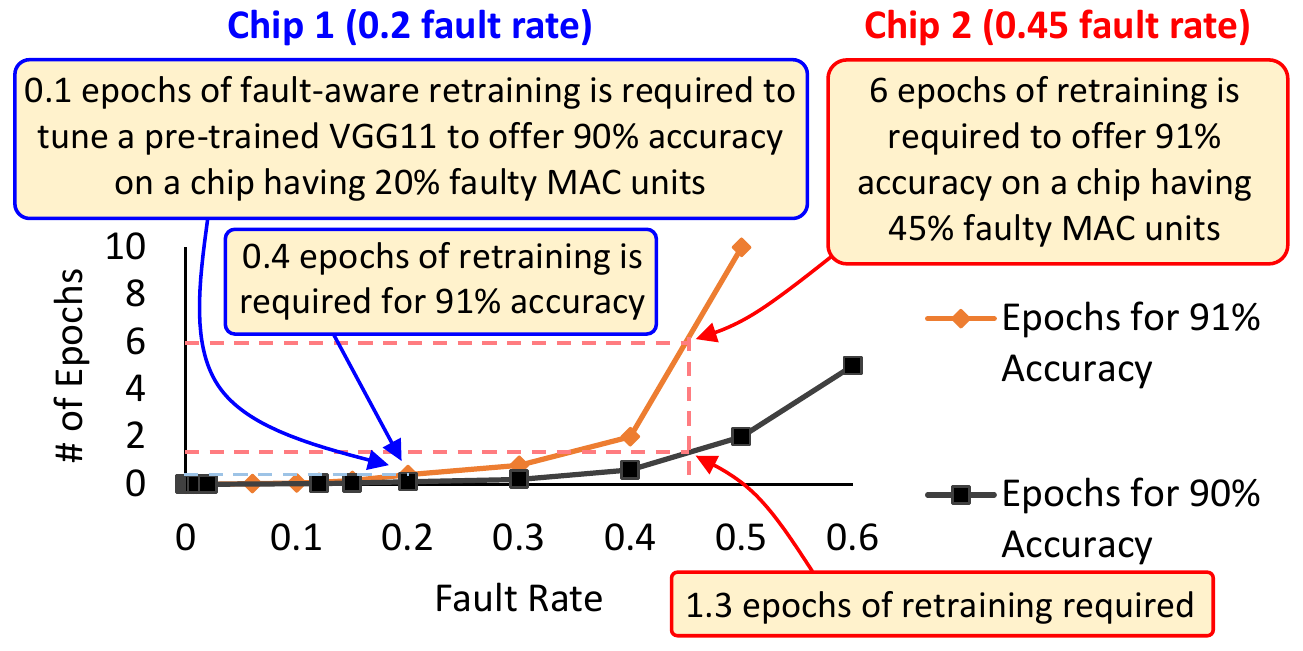}
\caption{An example showing how the amount of fault-aware retraining is computed for faulty chips. For this example, we considered two chips (1 and 2) having 0.2 and 0.45 fault rates (respectively).}
\label{fig:Example_Retraining_Amount_Selection}
\end{figure}

For computing the required amount of fault-aware retraining for a given faulty chip, we take the fault rate of the chip and check the corresponding amount of retraining required from the resilience plot. 
For fault rates at which retraining amount is not defined, we use bilinear interpolation using the nearest two data points to estimate the required amount of retraining. Fig.~\ref{fig:Example_Retraining_Amount_Selection} shows an example of how this is done for VGG11 (trained on Cifar10 dataset) for two different faulty chips having different fault rates. 

\subsection{Resilience-Driven Grouping and Fusion of Fault Maps (Step~3)}

Grouping multiple fault maps together can help in reducing the overall retraining cost required to tune a DNN for multiple faulty chips. However, as seen in Fig.~\ref{fig:Resilience_Results}a, the required amount of retraining usually increases exponentially with the increase in the fault rate. 
Therefore, there is a significant chance that combining multiple chips do not offer any benefit rather increases the overall retraining cost. For example, consider two faulty chips, A and B, having different number of faults. The fault rate (i.e,, the probability of a component being faulty) of each chip can be computed using: 

\begin{equation}
    Pr = \frac{\#\:of\:faulty\:components}{Total\:components}
\end{equation}

Now, assuming the fault rates of the chips are represented by $Pr_A$ and $Pr_B$ (respectively), and the faults are introduced in each chip randomly and independent of the faults in other chips, the fault rate of the combined fault map can be computed by:

\begin{equation}
    Pr_{combAB} = Pr_A + Pr_B - Pr_A \cap Pr_B
\end{equation}

Here, $Pr_{combAB}$ represents the combined fault rate and $Pr_A \cap Pr_B$ represents the probability of a component being faulty in both $A$ and $B$ chips. Given that the fault maps are independent of each other, $Pr_A \cap Pr_B =$ $Pr_A * Pr_B$. 
Specifically at low and moderate fault rates, the $Pr_A * Pr_B$ factor is negligible and $Pr_{combAB} \approx Pr_A + Pr_B$. 
Combining two fault maps is only beneficial if the retraining cost at $Pr_{combAB}$ is less than the sum of the retraining cost at $Pr_A$ and the retraining cost at $Pr_B$. Given the exponential nature of the resilience plots (see Fig.~\ref{fig:Resilience_Results}a), this is only possible when there is a significant correlation between the two fault maps, i.e., when $Pr_A \cap Pr_B$ is significant and larger than $Pr_A * Pr_B$. 
Moreover, as the rate of change of retraining cost is relatively smaller at lower fault rates compared to at higher fault rates, it seems beneficial to start searching for such pairs at lower fault rates and then progress towards higher fault rate chips. 
Therefore, based on the above two observations, we propose Algo.~\ref{Algo:2} for grouping and fusion of fault maps to reduce the overall fault-aware retraining cost. 

The algorithm mainly sorts the given fault maps in the increasing order of their fault rates, 
and then, for each fault map, starting from the fault map having the least fault rate, pairs it with (at maximum) $M$ randomly selected fault maps and selects the pair that offers the least fault rate after fusion. 
Then, the algorithm evaluates the savings, and if the fusion results in some reduction in the overall retraining cost, it removes the two fault maps from the list and adds the fused fault map at the location based on its fault rate. 
The algorithm also keeps track of these changes, as they define which fused fault map corresponds to which fault maps.

\begin{algorithm}[h]
\footnotesize
\caption{Grouping and fusion of fault maps}
\label{Algo:2}
\begin{algorithmic}[1]
\Statex \textbf{Inputs:} Fault maps (\textbf{\textit{FMs}}); Table of required amount of retraining at different fault rates (\textbf{\textit{TRC}}); Number of comparisons per fault map ($M$); Number of iterations ($K$)
\Statex \textbf{Outputs:} Merged fault maps (\textbf{\textit{MFMs}}); Table with links between \textbf{\textit{MFMs}} and \textbf{\textit{FMs}} (\textbf{\textit{T\_Link}})
\Statex \textbf{Initialize:} \textbf{\textit{MFMs}} = [ ]; \textbf{\textit{T\_Link}} = [[1], [2], ..., [number of fault maps in \textbf{\textit{FMs}}]]
\State \textbf{\textit{FRs}} = Compute fault rates of the given fault maps \textbf{\textit{FMs}}
\State \textbf{\textit{SFRs}} = Sort \textbf{\textit{FRs}} in ascending order
\State \textbf{\textit{SFMs}} = Arrange fault maps in \textbf{\textit{FMs}} using the corresponding indexes of \textbf{\textit{SFRs}}
\State \textbf{\textit{T\_Link}} = Arrange entities in \textbf{\textit{T\_Link}} using the corresponding indexes of \textbf{\textit{SFRs}}
\State \textbf{\textit{MFMs}} = \textbf{\textit{SFMs}}

\For{$j$ = 1:K}
\State $i=1$
\While{$i <$ number of fault maps in \textbf{\textit{MFMs}}}
\State $FM$ = Select $i^{th}$ fault map from \textbf{\textit{MFMs}}
\State \textbf{\textit{RFMs}} = Randomly select (at maximum) $M$ number of fault maps from \textbf{\textit{MFMs}}(:,:,$i+1$:end)
\State \textbf{\textit{R\_Savings}} = Compute relative amount of savings that would be achieved by combining $FM$ with each fault map in \textbf{\textit{RFMs}} individually. Use \textbf{\textit{TRC}} with bilinear interpolation to compute the retraining costs. 
\State [$Saving$, $I$] = min(\textbf{\textit{R\_Savings}})
\State $i$ $+= 1$
\If{$Saving > 0$}
\State Remove $FM$ and the $I^{th}$ fault map in \textbf{\textit{RFMs}} from \textbf{\textit{MFMs}}
\State Remove the corresponding fault rates from \textbf{\textit{SFRs}} as well
\State Add the merged fault map generated by combining $FM$ with the $I^{th}$ fault map in \textbf{\textit{RFMs}} in \textbf{\textit{MFMs}} at the location based on its fault rate
\State Add the fault rate of the merged fault map in \textbf{\textit{SFRs}}
\State Combine $i^{th}$ entity in \textbf{\textit{T\_Link}} with the entity corresponding to the $I^{th}$ fault map in \textbf{\textit{RFMs}} and place it back in \textbf{\textit{T\_Link}} at the location corresponding to the location of the merged fault map
\State $i$ $-= 1$
\EndIf
\EndWhile
\EndFor
\State \Return \textbf{\textit{MFMs}}, \textbf{\textit{T\_Link}}
\end{algorithmic}
\end{algorithm}

\color{black}
\section{Results and Discussion}
\label{Sec:results}

\subsection{Experimental Setup}
\label{sec:experimental_setup}

\begin{figure}[h]
\centering
\includegraphics[width=1\linewidth]{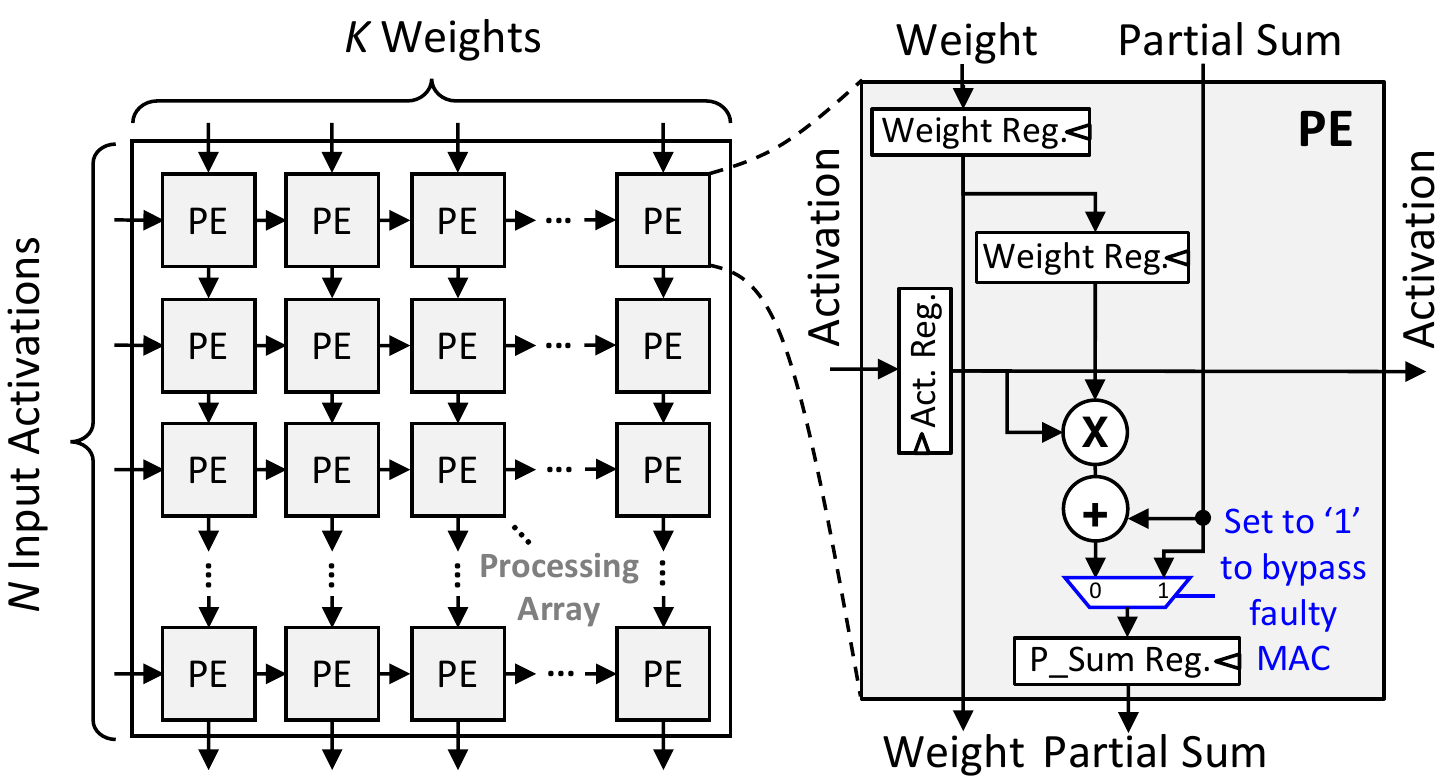}
\caption{Fault-tolerant systolic array design proposed in~\cite{zhang2018analyzing}.}
\label{fig:Mod_Hardware}
\end{figure}

To evaluate the effectiveness of the proposed technique for reducing the overall retraining cost, we consider the case of permanent faults induced due to imperfection in the manufacturing process in the computational array of a DNN hardware accelerator.  
For this study, we consider the architecture shown in Fig.~\ref{fig:hardware}. 
Further, we assume that the architecture is equipped with the additional circuitry required to support FAP~\cite{zhang2018analyzing}. 
Fig.~\ref{fig:Mod_Hardware} shows the modified systolic array design proposed in~\cite{zhang2018analyzing} that we use for this case study. 
For all the analysis, we assume the size of the systolic array to be $256 \times 256$. 

\begin{figure}[h]
\centering
\includegraphics[width=1\linewidth]{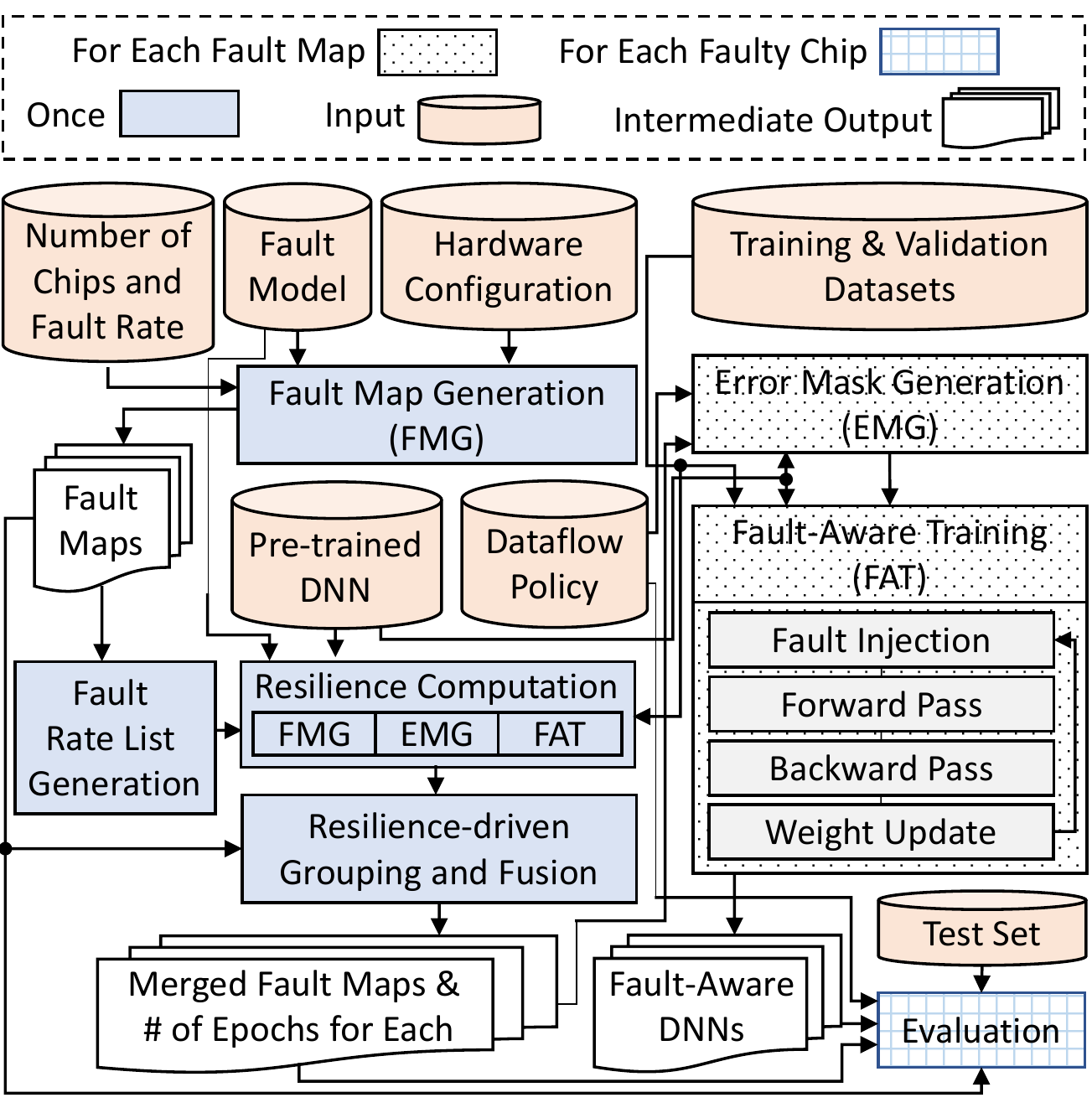}
\caption{Experimental Setup used for evaluating our proposed eFAT framework}
\label{fig:Exp_Setup}
\end{figure}

The complete experimental setup is illustrated in Fig.~\ref{fig:Exp_Setup}. 
We built our entire framework using Python and the PyTorch library. 
Our framework first accepts the number of chips together with fault rate, fault model and hardware configuration to generate fault maps. 
Note, similar to the earlier studies such as~\cite{zhang2018analyzing} and~\cite{abdullah2020salvagednn}, we considered a random fault injection model for generating fault maps in this work. 
Using the generated fault maps and Algo.~\ref{Algo:1}, the \textit{fault rate list generation} module then generates a list of fault rates for the resilience computation step. 
The \textit{resilience computation} module then generates the \textit{amount of retraining vs. fault rate} table, which is used in \textit{resilience-driven grouping and fusion} module to merge multiple fault maps and define the required amount of retraining for each chip/group of chips. 
The estimated values are then used to tune the given pre-trained DNN for the generated fault maps using FAT.
The generated set of fault-aware DNNs are then evaluated using test set to generate the final results. 

\subsection{Resilience Trends Across DNNs and Datasets}

\begin{figure}[ht]
\centering
\includegraphics[width=1\linewidth]{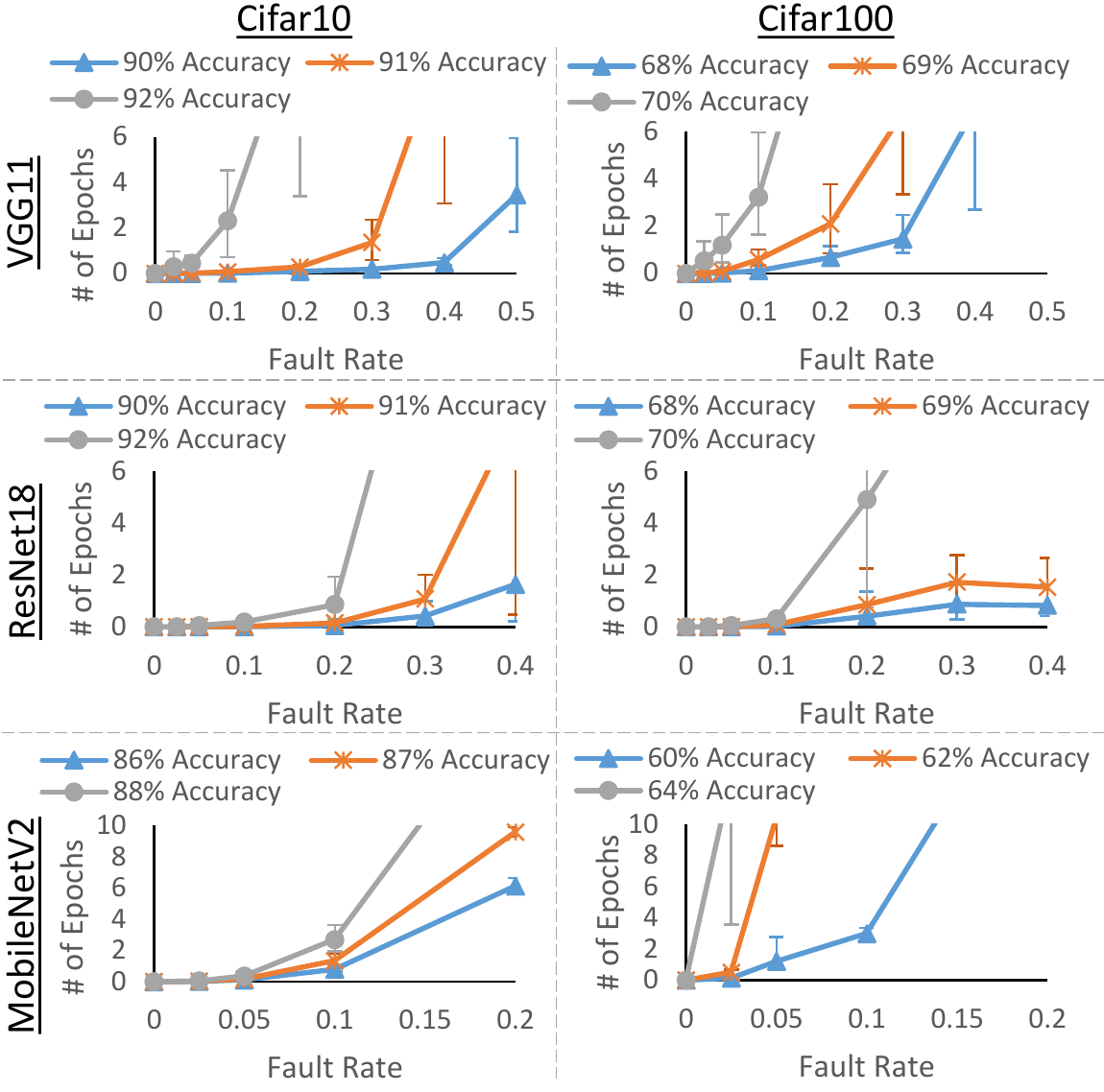}
\caption{Resilience of different DNN and dataset pairs.}
\label{fig:Resilience_plots}
\end{figure}

Fig.~\ref{fig:Resilience_plots} shows the resilience trends for different DNN and dataset pairs. 
We mainly considered three different DNNs, i.e., VGG11, ResNet18, and MobileNetV2, and two different dataset, i.e., Cifar10 and Cifar100, to demonstrate the impact of fault rate on the number of epochs required to reach the user-defined accuracy constraint. 
For each data point in the figure, we repeated the experiment five times and reported the minimum and maximum number of epochs as well along with the mean. 
The error bars show that the use of mean values can lead to under-training, while the use of maximum reported values will result in higher confidence that the resultant fault-aware DNN meets the user-defined accuracy constraint. 
Therefore, we propose to use the maximum bounds (wherever possible) to estimate the required amount of retraining for a given faulty chip. 

\subsection{Comparison with State of the Art}

\begin{figure*}[ht]
\centering
\includegraphics[width=0.9\linewidth]{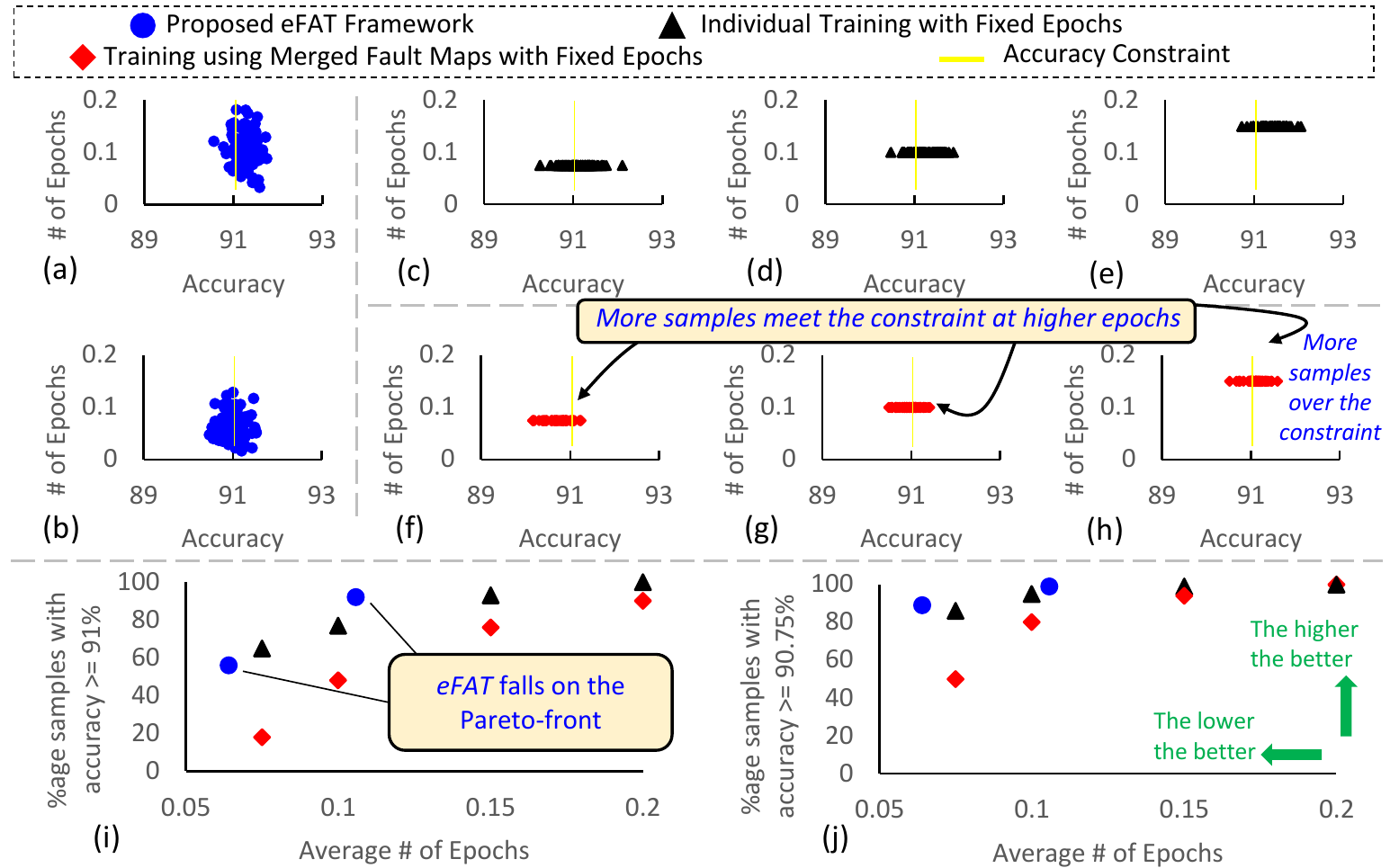}
\caption{Comparison with the state-of-the-art methods. (a) Results generated through proposed \textit{eFAT} framework using max. values from the resilience analysis for retraining amount estimation. (b) Results generated through \textit{eFAT} using mean values from resilience analysis for retraining amount estimation.
(c), (d) and (e) corresponds to the cases where VGG11 is trained for each faulty chip individually. (f), (g) and (h) corresponds to the cases where fault maps are paired and merged using the method proposed in~\cite{hoang2021tre}. (i) and (j) present the summary of the above plots and highlight the overall performance of eFAT in comparison to fixed policy fault-aware retraining and the method proposed in~\cite{hoang2021tre}}
\label{fig:SoA_comparison}
\end{figure*}

To highlight the effectiveness of the proposed technique, we compared our \textit{eFAT} framework with fixed-policy retraining methods proposed in~\cite{zhang2018analyzing} and~\cite{hoang2021tre}. 
Although~\cite{hoang2021tre} discussed the concept of merging multiple fault maps, it lacks a systematic methodology for exploiting the concept. 
Figs.~\ref{fig:SoA_comparison}a and~\ref{fig:SoA_comparison}b show the results of the proposed methodology when employed for retraining VGG11 (trained on the Cifar10 dataset) for 100 fault chips. 
For these experiments, we considered a setup similar to~\cite{zhang2018analyzing} and injected faults randomly in the computational array of a DNN accelerator. 
For defining the fault rates of the chips, we generated an array of fault rates sampled from a Gaussian distribution with mean = 0.1 and sigma = 0.02. 
Figs.~\ref{fig:SoA_comparison}c, \ref{fig:SoA_comparison}d and \ref{fig:SoA_comparison}e correspond to the cases where the DNN is trained for each faulty chip individually for a pre-specified number of epochs. 
These figures show that as the amount of retraining is increased the number of samples that meet the accuracy constraint increases. 
Figs.~\ref{fig:SoA_comparison}f, \ref{fig:SoA_comparison}g and \ref{fig:SoA_comparison}h correspond to~\cite{hoang2021tre}.
For this, considering the fact that retraining cost increases exponentially with fault rate, we only simulated the case where all the faulty chips are randomly divided into pairs and then each pair is merged into one fault map. 
Comparing Figs.~\ref{fig:SoA_comparison}c, \ref{fig:SoA_comparison}d and \ref{fig:SoA_comparison}e with Figs.~\ref{fig:SoA_comparison}f, \ref{fig:SoA_comparison}g and \ref{fig:SoA_comparison}h, it can be concluded that individual training performs better than randomly merging fault maps and then retraining. 
The results of Figs.~\ref{fig:SoA_comparison}a to \ref{fig:SoA_comparison}h are summarized in Figs.~\ref{fig:SoA_comparison}i and \ref{fig:SoA_comparison}j. 
The figures show that the proposed \textit{eFAT} framework produced more robust models with lesser training compared to the state-of-the-art techniques. 

\section{Conclusion}
\label{Sec:conclusion}

In this paper, we proposed, \textit{eFAT}, a framework for reducing the overheads of fault-aware retraining when used for tuning a given DNN for numerous faulty chips. We mainly addressed the following research questions. RQ1: How to compute the amount of retraining required for tuning the given DNN for a specific faulty chip. RQ2: Under which circumstances the fault maps of multiple faulty chips should be fused together to further reduce the overall retraining overheads? 
The results showed that the proposed technique can significantly reduce the overall retraining costs. 

\def\bibfont{\footnotesize}
\bibliographystyle{IEEEtran}
\bibliography{biblio}

\end{document}